\documentclass[twocolumn,showpacs,preprintnumbers,amsmath,amssymb,superscriptaddress]{revtex4}

\usepackage{amsmath,amsfonts,amssymb}
\usepackage[english]{babel} 
\usepackage[latin1]{inputenc} 
\usepackage[T1]{fontenc}
\usepackage{color}
\usepackage{float}
\usepackage{verbatim}
\usepackage{graphicx}

\begin{document}

\title{Mean cover time of one-dimensional persistent random walks}

\author{Marie Chupeau}
\affiliation{Laboratoire de Physique Th\'eorique de la Mati\`ere Condens\'ee (UMR CNRS 7600), Universit\'e Pierre et Marie Curie, 4 Place Jussieu, 75255
Paris Cedex France}

\date{\today}
\author{Olivier B\'enichou}
\affiliation{Laboratoire de Physique Th\'eorique de la Mati\`ere Condens\'ee (UMR CNRS 7600), Universit\'e Pierre et Marie Curie, 4 Place Jussieu, 75255
Paris Cedex France}

\author{Rapha\"el Voituriez}
\affiliation{Laboratoire de Physique Th\'eorique de la Mati\`ere Condens\'ee (UMR CNRS 7600), Universit\'e Pierre et Marie Curie, 4 Place Jussieu, 75255
Paris Cedex France}
\affiliation{Laboratoire Jean Perrin, FRE 3231 CNRS /UPMC, 4 Place Jussieu, 75255
Paris Cedex}

\begin{abstract}
The cover time is defined as the time needed for a random walker to visit every site of a confined domain. Here, we focus on persistent random walks, which provide a minimal model of random walks with short range  memory. We derive the exact expression of the mean cover time of a one-dimensional lattice by such a persistent random walk, both for  periodic and reflecting boundary conditions.
\end{abstract}

\maketitle

\section{Introduction}

How long does it take a random walker to visit every site of a confined domain?  
This time, known as the cover time in the mathematics literature, has important applications in the context of  robotics or computer science (for instance in protocol testing \cite{Mihail}). More generally, it is an alternative to first-passage times, which have been extensively studied in the last few years in fields as varied as chemical reaction kinetics \cite{Obenichou:2008} or animal behaviour \cite{Benichou:2011}. Indeed, the cover time can also be used to quantify the efficiency of the search processes where every site has to be visited.

However, exact results on cover times are scarce. Important steps were achieved in \cite{Yokoi}, where the mean cover time of an interval was analytically calculated for one dimensional symmetric nearest-neighbour random walks, both for periodic and reflecting boundary conditions. In  dimensions greater or equal to three, Aldous \cite{Aldous88} has determined the  leading behaviour of the mean cover-time in the limit of large domain size, which was reproduced by numerical simulations in \cite{Nemirovsky}. In the physics literature, Hilhorst and Brummelhuis have extended these results to the two-dimensional case in \cite{Hilhorst91}, which have been since then refined in the mathematics literature (for instance in \cite{Dembo,Ding}). 

All these results were obtained  in the case of symmetric nearest-neighbour random walks. Here, we focus on persistent random walks, which provide a minimal model of random walks with short range  memory  and have proved to play an important role in various contexts, including search processes \cite{Ernst,Benichou:2011,Tejedor:2012}. In this paper, based on rather elementary methods, we derive the exact expression of the mean cover time of a one-dimensional lattice by a persistent random walk with periodic and  reflecting boundary conditions, which quite surprisingly has not been considered so far to the best of our knowledge.

\section{Periodic boundary conditions}

\vspace{0.5cm}

\begin{figure}[h]
\centering
\includegraphics[width=250pt]{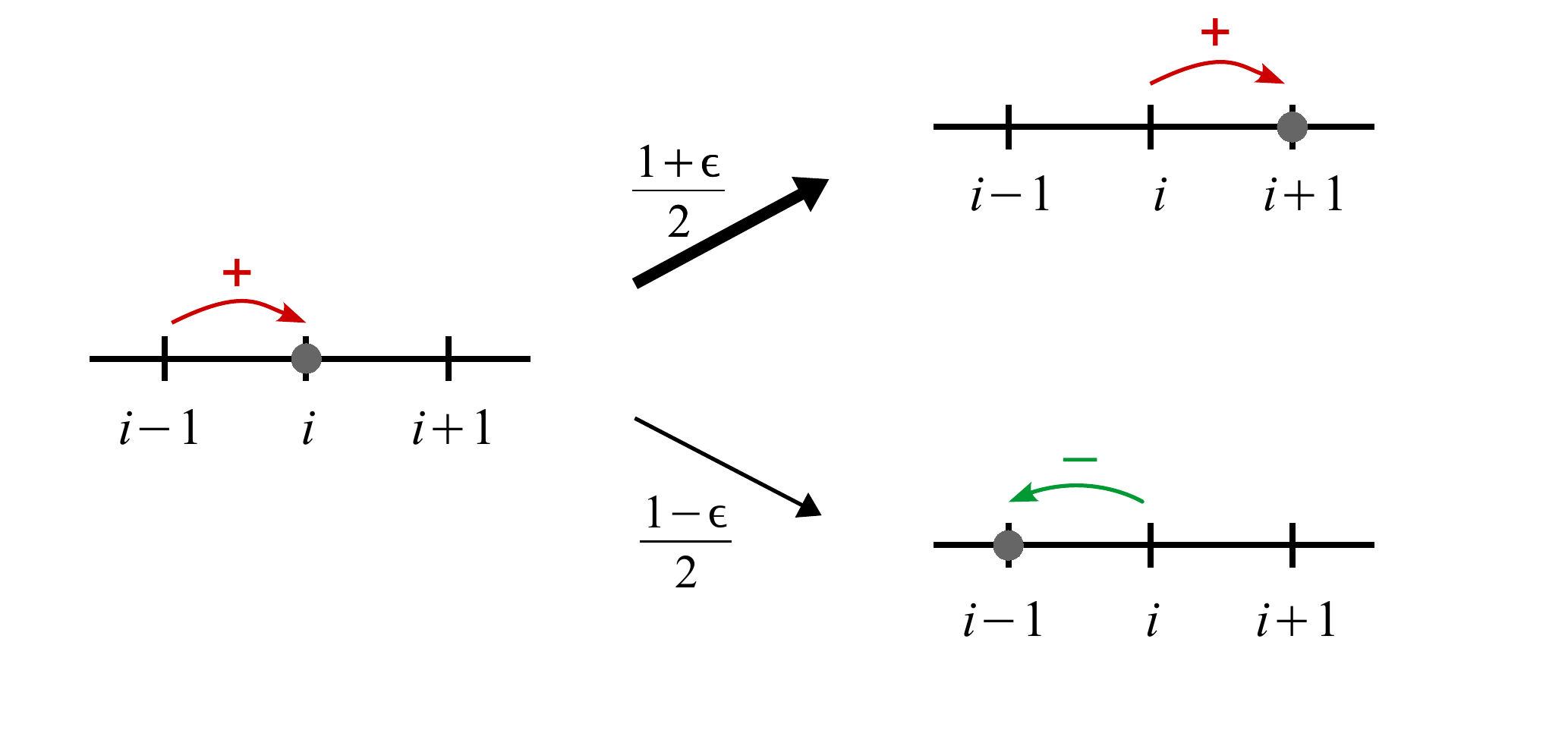}
\caption{Illustration of a step in persistent walk. If the previous step was made rightwards, the probability of making another step to the right is higher.}
\label{persist}
\end{figure}

We consider a discrete time persistent random walk  on a discrete one dimensional lattice (see Fig.\ref{persist}). At each time step, the random walker has a probability  $\dfrac{1+\epsilon}{2}$ to continue in the same direction as the previous step  and  $\dfrac{1-\epsilon}{2}$ to go backward. Note that the cases $\epsilon=1$ and $\epsilon=-1$ are both particular because they are deterministic when the first step is performed. The case $\epsilon=1$ is a purely ballistic walk and $\epsilon=-1$ is a back-and-forth motion between two adjacent sites leading to trapping effects and thus diverging cover times. In the following, we take $-1< \epsilon <1$

We first consider the case of an interval of $N$ sites with periodic boundary conditions,  in which  all sites are equivalent. In particular, the mean cover time $\tau(N)$ defined as the mean time needed to visit all the sites of the interval $[0,...,N-1]$ does not depend of the starting point in this case. 
Following Yokoi \textit{et al.} \cite{Yokoi},  we write the mean cover time for a $N$ site ring $\tau(N)$ as  the sum of the mean cover time for $N-1$ sites among $N$ and the mean time $\bar{t}(N)$ to visit the last site:
\begin{equation}
\tau(N)=\tau(N-1)+\bar{t}(N).
\label{recurrence_periodic}
\end{equation}
 Here we have used that the mean cover time for $k$ consecutive sites of a ring of $N$ sites ($k\le N$) is given by $\tau(k)$.

We next introduce $T_+(d)$ (resp. $T_-(d)$) as the mean time needed   to reach $x_0+d$ for the first time,  knowing that the walker arrived at site $x_0$ from site $x_0-1$ (resp. from site $x_0+1$) at time step 0. These quantities also depend on the number of sites $N$, but for the sake of clarity, this dependency is not explicitly written. Note that, because of the boundary conditions, these conditions do not depend of $x_0$.  It is then easily checked that  $\bar{t}(N)=T_+(1)$, which we now calculate. 

To evaluate $\bar{t}$, we actually compute the function $T_+(d)$ for all $d$, which is solution for $d\ge1$ of the coupled backward equations:
\begin{equation}
\begin{cases}
T_+(d)=\dfrac{1+\epsilon}{2} \; T_+(d-1)+\dfrac{1-\epsilon}{2} \; T_-(d+1)+1 \\[.9em]
T_-(d)=\dfrac{1-\epsilon}{2} \; T_+(d-1)+\dfrac{1+\epsilon}{2} \; T_-(d+1)+1
\end{cases}
\end{equation}
obtained by partitioning over the first step of the walk, complemented by the boundary condition $T_{\pm}(0)=0$.
Combining these two equations yields:
\begin{equation}
\label{rec1}
T_+(d)-2 \,T_+(d-1)+T_+(d-2)+2 \,\dfrac{1-\epsilon}{1+\epsilon}=0
\end{equation}
which is solved by
\begin{equation}
T_+(d)=\lambda+\mu d  -\dfrac{1-\epsilon}{1+\epsilon}d^2
\end{equation}
The boundary condition $T_+(0)=0$ gives $\lambda=0$ while  $\mu$ is determined by making use of the periodicity of the ring: $T_+(d)=T_-(N-d)$. This finally gives
\begin{equation}
T_+(d)=d \;  \left( \dfrac{\dfrac{1-\epsilon}{1+\epsilon}\left(1-\dfrac{(1-\epsilon)(N-2)^2}{2} \right)+1}{1-\dfrac{(1-\epsilon)(N-2)}{2}}-\dfrac{1-\epsilon}{1+\epsilon} \; d \right),
\end{equation}
and in particular
\begin{equation}
\bar{t}(N)=T_+(1)=\dfrac{1-\epsilon}{1+\epsilon} \; N + \dfrac{3 \epsilon-1}{1+\epsilon}.
\end{equation}
Using Equation (\ref{recurrence_periodic}) iteratively and noting that $\tau(2)=1$, we get
\begin{eqnarray}
&\tau(N)& =\sum\limits_{i=3}^N \bar{t}(i) + \tau(2) \nonumber \\ 
 &  &= \dfrac{(N+3)(N+2)}{2} \dfrac{1-\epsilon}{1+\epsilon} + (N-2)\dfrac{3 \epsilon-1}{1+\epsilon}+1 \nonumber \\
\end{eqnarray}

It is then found that Equation (\ref{recurrence_periodic}) is solved by
\begin{equation}
\tau(N)=\dfrac{1-\epsilon}{2 \, (1+\epsilon)} \, N^2+ \dfrac{5 \epsilon-1}{2 \, (1+\epsilon)} \, N - \dfrac{2 \epsilon}{1+\epsilon}
\label{periodic}
\end{equation}
which provides an exact explicit solution of the mean cover time.

Note that in the large $N$ limit, the mean cover time $N^2/4D$ of a regular random walk with diffusion coefficient $D$ is recovered, where 
\begin{equation}
D=\frac{1+\epsilon}{2 \, (1-\epsilon)}
\end{equation}
 can be shown to be the diffusion coefficient  of a one-dimensional persistent random walk 
 defined in the large time limit.
 
 In contrast, if the persistence length $l_p\equiv\frac{1}{1-\epsilon}$ is of order $N$, one finds that the mean cover time scales linearly with $N$, as expected for a ballistic-like motion.

\section{Reflecting boundary conditions}

\begin{figure}[h] 
\centering
\includegraphics[width=150pt]{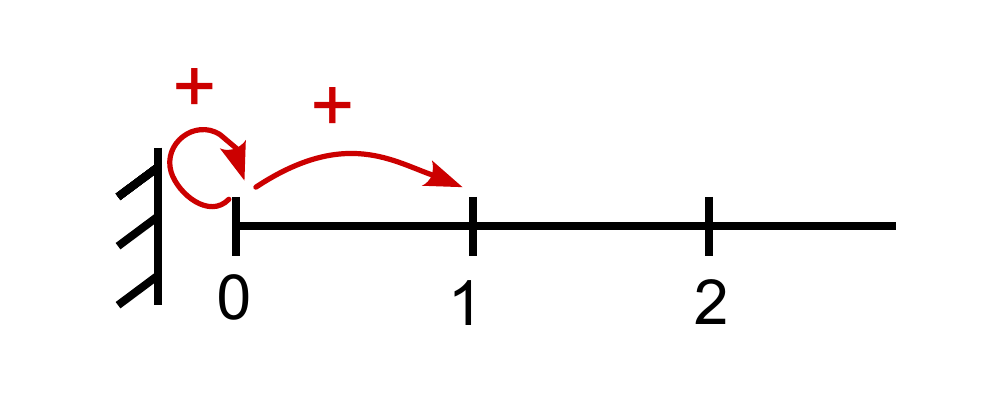}
\caption{After reflecting on the edge, we take the convention that the walker arrives at site $0$ from the left.}
\label{CLR}
\end{figure}

We now focus on the case of a chain of $N$ sites, labelled from $0$ to $N-1$, with reflecting boundary conditions (see Fig.\ref{CLR}).  To compute the cover time, we  split the process of covering the whole domain  in two steps:
(i) the walker first reaches one of the edges of the domain ($0$ or $N-1$); 
(ii) the walker then needs to cross the domain to reach the other edge.

\subsection{Reaching the first edge}

In this section, we  introduce two  quantities: the splitting probability $\Pi_{z}(y|x)$, defined as the probability of reaching the point $y$ before the point $z$ starting from the point $x$, and the associated mean conditional time $T_z(y|x)$, defined as  the mean time needed to reach the point $y$  starting from the point $x$, knowing that $z$ has remained unvisited. 
The mean time needed to reach any of the two edges is then given by 
\begin{equation} \label{keyeq}
T_{edge}=\Pi_{N-1}(0|x) T_{N-1}(0|x) + \Pi_{0}(N-1|x) T_{0}(N-1|x).
\end{equation}
We now calculate successively the splitting probabilities and the mean conditional times  involved in this equation. As before, these two quantities depend on the direction of the walker when it arrives at site $x$ at step zero. 

\subsubsection{Splitting probabilities}
The splitting probabilities $\Pi_{N-1}^+(0|x)$ and $\Pi_{N-1}^-(0|x)$ 
are related to $\Pi_{N-1}(0|x)$
by 
\begin{equation} 
\Pi_{N-1}(0|x)= \frac{1}{2} \, \Pi_{N-1}^+(0|x) + \frac{1}{2} \, \Pi_{N-1}^-(0|x).
\end{equation} 
They satisfy the  two coupled backward equations for $x\in\{1,..,N-2\}$:
\begin{equation} \label{pi}
\begin{cases}
\Pi_{N-1}^+(0|x)=\dfrac{1+\epsilon}{2} \; \Pi_{N-1}^+(0|x+1)+\dfrac{1-\epsilon}{2} \; \Pi_{N-1}^-(0|x-1) \\[.9em]
\Pi_{N-1}^-(0|x)=\dfrac{1+\epsilon}{2} \; \Pi_{N-1}^-(0|x-1)+\dfrac{1-\epsilon}{2} \; \Pi_{N-1}^+(0|x+1) 
\end{cases}
\end{equation}

\vspace*{0.5cm}
The two splitting probabilities are easily seen to be solutions of the second-order recurrence equation  
\begin{equation}
\Pi_{N-1}^{\pm}(0|x+1)-2\, \Pi_{N-1}^{\pm}(0|x)+\Pi_{N-1}^{\pm}(0|x-1)=0 
\end{equation}
valid for $x\in\{2,..,N-2\}$ for  $\Pi_{N-1}^{+}(0|x)$ and for  $x\in\{1,..,N-3\}$ for $\Pi_{N-1}^{-}(0|x)$.
The general solutions read
\begin{equation}
\begin{cases}
\Pi_{N-1}^+(0|x)=\lambda + \mu x & \textrm{for}\; x\in\{1,..,N-1\} \\
\Pi_{N-1}^-(0|x)=\lambda' + \mu' x & \textrm{for}\; x\in\{0,..,N-2\} 
\end{cases}
\end{equation}
Using Eq.(\ref{pi}) and the following boundary conditions
\begin{equation}
\begin{cases}
\Pi_{N-1}^+(0|N-1)=0 \\
\Pi_{N-1}^-(0|0)=1,
\end{cases}
\end{equation}
we finally obtain:
\begin{equation}
\begin{cases}
\Pi_{N-1}^+(0|x)= \dfrac{1-\epsilon}{N(\epsilon-1)+1-3 \epsilon} \; (x-N+1) \\[.9em]
\Pi_{N-1}^-(0|x)=1 + \dfrac{1-\epsilon}{N(\epsilon-1)+1-3 \epsilon} \; x.
\end{cases}
\end{equation}

\subsubsection{Mean conditional time} 
Defining
\begin{equation}
\begin{cases}
R^+_{N-1}(0|x)=\Pi_{N-1}^+(0|x) \; T_{N-1}^+(0|x) \\
R^-_{N-1}(0|x)=\Pi_{N-1}^-(0|x) \; T_{N-1}^-(0|x) \\
 R_{N-1}(0|x)=\frac{1}{2} \; R^+_{N-1}(0|x) + \frac{1}{2} \; R^-_{N-1}(0|x)
 \end{cases}
 \end{equation}
 it is seen following \cite{Kampen:1992,redner} that for $x\in\{1,..,N-1\}$
 \begin{eqnarray}
\label{backwardR}
&\dfrac{1+\epsilon}{2} \;R^+_{N-1}(0|x+1) + \dfrac{1-\epsilon}{2} \; R^-_{N-1}(0|x-1)-R^+_{N-1}(0|x)\nonumber\\
&= - \Pi_{N-1}^+(0|x)
\end{eqnarray}
\noindent and similarly, for $x\in\{0,..,N-2\}$
\begin{eqnarray}
&\dfrac{1+\epsilon}{2} \; R^-_{N-1}(0|x-1) + \dfrac{1-\epsilon}{2} \; R^+_{N-1}(0|x+1) -R^-_{N-1}(0|x) \nonumber\\
&=- \Pi_{N-1}^-(0|x)
\end{eqnarray}

\vspace*{0.5cm}
We again combine these two equations to obtain a recurrence equation on $R^+_{N-1}(0|x)$ for $x\in\{1,..,N-3\}$:

\begin{eqnarray}
\label{recR}
&&R^+_{N-1}(0|x+2)-2 \, R^+_{N-1}(0|x+1)+R^+_{N-1}(0|x)=\nonumber\\
&-& \dfrac{2}{1+\epsilon} \; \Pi_{N-1}^+(0|x+1)
-\dfrac{1-\epsilon}{1+\epsilon} \; \Pi_{N-1}^-(0|x)+\Pi_{N-1}^+(0|x),\nonumber\\
\end{eqnarray}
whose solution is given, for $x\in\{1,..,N-1\}$ by 
\begin{equation}
R^+_{N-1}(0|x)=\lambda+\mu x + \alpha x^2 + \beta x^3
\end{equation}

with
\begin{equation}
\begin{cases}
\begin{split}
\alpha=\dfrac{1-\epsilon}{1+\epsilon} \; \dfrac{\left(1-\epsilon\right)\left(N-1\right)}{N \left(\epsilon-1\right)+1-3 \epsilon} \\[1.1em]
\beta=\dfrac{1-\epsilon}{3 \left(1+\epsilon\right)} \; \dfrac{1-\epsilon}{N \left(1-\epsilon\right)+3 \epsilon-1} 
\end{split}
\end{cases}
\end{equation}
The constants $\lambda$ and $\mu$ are determined by using  the boundary condition 
\begin{equation}
R^+_{N-1}(0|N-1)=0
\end{equation}
and writing Eq. (\ref{backwardR}) at $x=1$:
\begin{equation}
R^+_{N-1}(0|1)=\dfrac{1+\epsilon}{2} \; R^+_{N-1}(0|2)+\Pi_{N-1}^+(0|1).
\end{equation}

This finally leads to 
\begin{eqnarray}
&&R^+_{N-1}(0|x)=\left(x-N+1 \right) \times\nonumber\\\
&\times&\left(\mu +\alpha \left(x+N-1\right) + \beta \left(x^2 +\left(N-1 \right) x + \left(N-1 \right)^2 \right) \right)\nonumber\\
\end{eqnarray}
with 
\begin{eqnarray}
\mu&=&-\dfrac{2}{3}  \dfrac{\left(1-\epsilon\right)^2}{\left(1+\epsilon\right) \left(N \left(\epsilon-1\right)+1-3 \epsilon\right)^2} \nonumber\\
 &\times& \left[ \left(1+\epsilon\right) \left(3-N\right) \left(-N^2+3\right)-3 \; \dfrac{1+\epsilon}{1-\epsilon} \left(2-N\right) \right.\nonumber\\
&& \left. + \left(N-2\right) \left(-2 N^2+2 N+1\right) \right. \bigg]
\end{eqnarray}
We next obtain from Eq. (\ref{backwardR}) the expression of $R_-^0(x)$:
\begin{eqnarray}
&R^-_{N-1}(0|x)= \left( \mu + \dfrac{12 \epsilon}{1-\epsilon} \; \beta \left( \dfrac{1}{1-\epsilon} + N-2 \right) \right) x \nonumber\\
&- \dfrac{1-\epsilon}{1+\epsilon} x^2 +\beta x^3 
\end{eqnarray}

\vspace*{0.5cm}
Making finally use of
\begin{eqnarray}
R_0(N-1|x)&\equiv& \frac{1}{2} \, \Pi_{0}^+(N-1|x) \; T_{0}^+(N-1|x)\nonumber\\
&+&\frac{1}{2} \, \Pi_{0}^-(N-1|x) \; T_{0}^-(N-1|x)\nonumber\\
&=&R_{N-1}(0|N-1-x),
\end{eqnarray}
 we obtain:
\begin{eqnarray}
T_{edge}&=&R_{N-1}(0|x)+R_0(N-1|x)\nonumber\\
&=&-\dfrac{1}{1+\epsilon} \left[ \left(N-1\right) \left(\epsilon x-\epsilon -x \right) + x^2 \left(1-\epsilon\right) \right]\nonumber\\
\label{reflecting}
\end{eqnarray}

\subsection{Going from one edge to the other}

We first introduce $T_{\pm}(x)$, the mean time needed to go from $x$ to $x+1$, knowing the initial step $\pm$.
It satisfies:
\begin{align} \label{coupled}
T_+(x) &= \dfrac{1+\epsilon}{2} + \dfrac{1-\epsilon}{2} \big(1+T_-(x-1) +T_+(x) \big) \\[.9em]
T_-(x) &= \dfrac{1-\epsilon}{2} + \dfrac{1+\epsilon}{2} \big(1+T_-(x-1)+T_+(x) \big)
\end{align}

This leads to
\begin{equation}
T_+(x+1)=T_+(x)+2 \; \dfrac{1-\epsilon}{1+\epsilon}
\end{equation}
and finally yields for $x\in\{1,..,N-1\}$
 \begin{equation}
T_+(x)=2 \,  \dfrac{1-\epsilon}{1+\epsilon} \, x + T_+(0).
\end{equation}

In addition, 
\begin{equation}
T_+(0)= \dfrac{1+\epsilon}{2} + \dfrac{1-\epsilon}{2} \left(1+T_+(0) \right)
\end{equation}
so that finally
\begin{equation}
T_+(x)=2 \, \dfrac{1-\epsilon}{1+\epsilon} \, x + \dfrac{2}{1+\epsilon}
\end{equation}
 and straightforwardly, using Eq. (\ref{coupled}),
\begin{equation}
T_-(x)=2 \, (x+1)
\end{equation}
Finally, the mean time to cross the interval is given by
\begin{eqnarray}
T_{cross}&=&T_-(0)+T_+(1)+...+T_+(N-2)\nonumber \\
              &=&2+\dfrac{1-\epsilon}{1+\epsilon} \left(N-1\right) \left(N-2\right)+\dfrac{2}{1+\epsilon} \left(N-2\right) \nonumber\\
\end{eqnarray}

\subsection{Cover time}

We eventually obtain the cover time when the walker starts from a site $x$ that is not one of the edges of the domain:
\begin{eqnarray}
\tau_R(x)&=&T_{edge}+T_{cross}\nonumber\\
           &=&  -\dfrac{1}{1+\epsilon} \; \Big[ \left(N-1\right) \left(\epsilon x -\epsilon-x\right)+x^2 \left(1-\epsilon \right) \Big] \nonumber\\&+& \dfrac{1-\epsilon}{1+\epsilon} \left(N-1\right) \left(N-2\right) 
           + \dfrac{2}{1+\epsilon}\left(N-2\right)+2 \nonumber\\
            \label{ref2}
\end{eqnarray}
The cover time starting from one edge can also be written, partitioning on the first step:
\begin{eqnarray}
\tau_R(0) &=&\tau_R(N-1)\nonumber\\
&=&\dfrac{1}{2} \; \Big[ T_+(N-1|0)+T_+(N-1|1) \Big] +1\nonumber\\
            &=& \dfrac{1}{2} \; T_+(0)+T_+(1)+...+T_+(N-2)+1\nonumber\\
            &=&\dfrac{1}{1+\epsilon}+\dfrac{1-\epsilon}{1+\epsilon} \left(N-1\right) \left(N-2 \right)+\dfrac{2}{1+\epsilon} \left(N-2 \right)+1\nonumber\\
            &=&  \dfrac{1-\epsilon}{1+\epsilon}\left(N-1 \right) \left(N-2 \right) +\dfrac{2}{1+\epsilon} \left(N-2 \right)+\dfrac{2+\epsilon}{1+\epsilon}\nonumber\\
\end{eqnarray}

\section{Discussion and conclusion}

In this paper, we obtained the exact expression of the mean cover time in 1D for a persistent random walk with both periodic and reflecting boundary conditions. 
We  now examine how the expressions we found in periodic and reflecting cases behave when $\epsilon=0$ and $\epsilon=1$, and discuss the impact of the boundary conditions.\\

\subsection{Non-persistent case $\epsilon=0$}

It is the case of a usual Brownian walk. From Eqs.(\ref{periodic}) and (\ref{ref2}), we recover the  results of \cite{Yokoi}
 \begin{align}
&\tau_P=\dfrac{N(N-1)}{2}\\
&\tau_R(x)=N \, (N-1)+x \, (N-1-x)
\end{align}

\subsection{Ballistic walk $\epsilon=1$}

The probability of going on in the same direction at each step is $1$. Trajectories are then ballistic. For periodic boundary conditions, the result is obvious. For reflecting boundary conditions, the cover time is the average between the two only possible ways (starting left or right):
\begin{align}
\tau_P&=N-1\\
\tau_R&=\dfrac{1}{2} \, (N-1) + N=\dfrac{3}{2} \, N-\dfrac{1}{2}
\end{align}
This is recovered by  Eqs.(\ref{periodic}) and (\ref{ref2}).

\subsection{Impact of the boundary conditions on the cover time}

Let us compare the scaling of the cover time with reflecting boundary conditions and periodic boundary conditions when $N \to +\infty $, for fixed values of $x$ and $\epsilon$ (still with $-1<\epsilon<1$). One finds:
\begin{equation}
\dfrac{\tau_R}{\tau_P} = 2 + 2 \left( x-\dfrac{\epsilon}{1-\epsilon} \right) \dfrac{1}{N} + o\left( \dfrac{1}{N} \right)
\end{equation}

To leading order, it takes twice as much time in reflecting boundary conditions to cover the whole domain as in periodic boundary conditions. Notably, this result does not depend on the initial position $x$ nor on $\epsilon$. A monotonic dependence on these two parameters appears in the first-order term.

\vspace{1cm}

A first extension of this work concerns the determination of higher order moments of the cover time. Even if less straightforward, the methodology used in this article holds in principle. For instance, in the reflecting case, the decomposition used in the key equation (\ref{keyeq}) can still be used to relate moments of the cover time to the moments of the conditional first-passage times, which in turn should be analytically calculable.\\

In the context of the search processes mentioned in introduction, an important extension is the generalisation to higher space dimensions, and in particular to two dimensions. Such results would offer a new tool to quantify the efficiency of search processes, alternatively to the now classical first-passage times. In particular, an important question would be to determine whether the minimisation of the mean first-passage time found for persistent random walks in two dimensions in \cite{Tejedor:2012} still holds for the mean cover time, and if so, whether it is reached at the same persistence length.

\vspace*{1cm}

{\bf Acknowledgment} Support from European Research Council starting Grant No. FPTOpt-277998 is acknowledged.

\bibliographystyle{ieeetr}

\end{document}